\documentstyle[12pt,epsf,aps]{revtex}
\begin{document}
\widetext
\draft
\tighten
\newcommand{\E}{{\rm e}}
\newcommand{\fett}[1]{{\mbox{\boldmath$#1$}}}

\title{
Estimates of production rates of SUSY particles in ultra-relativistic
heavy-ion collisions}

\author{{\sc
M W Beinker$^1$,
B K\"ampfer$^2$,
G Soff$^1$} \\[3mm]}
\address{
$^1$Institut f\"ur Theoretische Physik, Technische Universit\"at Dresden,\\
01062~Dresden, Germany\\[1mm]
$^2$Research Center Rossendorf Inc.,
PF 510119, 01314 Dresden, Germany\\[1cm]}

\maketitle

\begin{abstract}
We estimate the production rates of supersymmetric particles in central
heavy-ion collisions at LHC.
The parton cascade model is used to seek for possible collective
phenomena which enlarge the production probability of very
heavy particles. Even if there is some indication of such cooperative
effects, higher energy and higher luminosity of proton beams
at LHC disfavor heavy-ion reactions in the search for supersymmetric
particles.
\end{abstract}


\section{Introduction}

The Standard Model (SM) of elementary particle physics is
experimentally verified to a large extent;
only the existence of the Higgs
boson remains still to be validated.
On the other hand, the SM does not provide a fundamental
unified description of the world since it contains too many and
unexplained parameters. Among the major tasks
of the future Large Hadron Collider (LHC) at CERN
there are attempts to find experimental evidence for the
Higgs boson and also to look for new phenomena indicating physics
beyond the SM. One prominent candidate for such an
extension of the SM is the Minimal Supersymmetric Standard Model
(cf.~\cite{Soff} for a recent survey).
The search for supersymmetric (SUSY) particles in proton-proton
collisions with beam energies of 7 GeV is therefore an important
project at the LHC.

Besides proton-proton collisions there is also planned to collide
$\vphantom{N}^{208}\text{Pb}$ with $\vphantom{N}^{208}\text{Pb}$
with beam energies of 
2.76 TeV per nucleon.
Here one is predominantly interested in the Quark Gluon Plasma
which is rather likely to be formed in such high-energy collisions.
If the lightest
strongly interacting SUSY particle has a rest mass below
1 TeV \cite{nilles:84},
the available energy in lead-lead collisions should also
be high enough to produce such supersymmetric particles.
The lower energies per nucleon and the
smaller luminosity for lead beams, compared to proton beams, seem
to reduce the possibility to find SUSY particles in
lead-lead collisions in relation to proton-proton collisions.
However, cooperative effects in heavy-ion collisions could enlarge
the rates of SUSY particles. Therefore one should investigate whether
collective effects can enhance the production of very heavy particles near or
slightly above the threshold. Such effects have proven important
in intermediate-energy heavy-ion collisions in the so-called
subthreshold production of hadrons \cite{sub_thresh}.

In this paper, we examine the possibility of collective effects and
estimate the expected production rates for supersymmetric particles in
central collisions of heavy nuclei at LHC energies.
Our calculations are based on the parton cascade model (PCM)
\cite{klausMueller,klaus_all,klausKapusta,klaus_rep}.

Our paper is organized as follows. In Section 2 we recall the basic
features of the PCM. The cross sections for the production of
SUSY particles in proton-proton collisions are presented in Section 3.
Section 4 contains the PCM results for lead-lead collisions at LHC.
Our conclusions can be found in Section 5.


\section{The model}

The PCM is well documented in several publications
\cite{klausMueller,klaus_all,klausKapusta,klaus_rep},
therefore, here we would like
to recall only the basic features. The PCM is a model which describes
nuclear collisions on the parton level. In the initialization stage,
the nuclei are resolved into their constituent partons
at a certain low-energy scale $Q_0$. The subsequent kinetics of the
two colliding bunches of partons is described by a set
of coupled quasi-classical transport equations of the Boltzmann type
\begin{eqnarray}
\label{transport}
p^\mu\partial_\mu F_a(p,r)&=&
\sum_k I_a^{k}(p,r),
\end{eqnarray}
where $F_a(p,r)$ is the phase-space density of partons of flavor $a$
at momentum $p$ and position $r$.
The left hand side of this equation describes the propagation.
On the right hand side,
$I_a^{k}(p,r)$ describes the interaction kernel where partons of flavor $a$
are involved. Possible interactions are elementary processes between
two partons in first order of
perturbative Quantum Chromo Dynamics (QCD)
and parton emission or absorption processes.
The explicit form of $I_a^{k}(p,r)$ can be expressed as
\begin{eqnarray}
\sum_k I_a^{k}(p,r)&=&
\sum_{b,c,d} J_{abcd}(p_a,r)+\sum_{bc} K_{abc}(p_a,r)\;,
\end{eqnarray}
where
\begin{eqnarray}
\label{kern4}
J_{abcd}(p_a,r) & = &
-\frac12\frac{1}{S_{ab}}\frac{1}{S_{cd}}
\int\frac{d^3p_b}{(2\pi)^3 2E_b}\,
\int\frac{d^3p_c}{(2\pi)^3 2E_c}\,
\int\frac{d^3p_d}{(2\pi)^3 2E_d}\,
(2\pi)^4\,\delta^{(4)}(p_1+p_b-p_c-p_d)
\nonumber \\[2mm]
& \times &
\left( F_a F_b \left[1\pm F_c\right]
\left[1\pm F_d\right]
\left|{\cal M}_{ab\rightarrow cd}\right|^2_{\rm eff}
- F_c F_d \left[1\pm F_a\right]
\left[ 1\pm F_b\right]
\left|{\cal M}_{cd\rightarrow ab}\right|^2_{\rm eff} \right)
\end{eqnarray}
and
\begin{eqnarray}
\label{kern3}
K_{abc}(p_a,r)&=&
-\frac12\frac1{S_{ab}}
\int\frac{d^3p_b}{(2\pi)^3 2E_b}\,
\int\frac{d^3p_c}{(2\pi)^3 2E_c}
(2\pi)^4\,\delta^{(4)}(p_1+p_b-p_c-p_d)
\nonumber \\[2mm]
& \times & \left( F_a F_b \left[1\pm F_c\right]
\left|{\cal M}_{ab\rightarrow c}\right|^2_{\rm eff}
- F_c \left[1\pm F_a\right]
\left[ 1\pm F_b\right]
\left|{\cal M}_{c\rightarrow ab}\right|^2_{\rm eff}\right)
\end{eqnarray}
(we use units with $\hbar = c =1$).
The factors $S_{ab}\equiv 1+\delta_{ab}$ serve for
the correct statistics in case of identical particles.
The effective matrix elements
$\left|{\cal M}_{\cdots}\right|^2_{\rm eff}$
are weighted with the particle densities $F_{\cdots}$ of the
incoming particles and are integrated over their momenta.
The factors of the form $\left[1 \pm F_{\cdots} \right]$ describe either Bose
enhancement (with $+$ sign) or Fermi suppression (with $-$ sign)
of the outgoing particles, respectively.

The effective matrix elements are corrected afterwards with respect to
the virtuality of the partons. They exhibit the form, e.g., for the process
$a b \to c d$
\begin{eqnarray}
\left|{\cal M}_{ab\rightarrow cd}\right|^2_{\rm eff}&=&
S_a(p_a;Q^2,Q^2_0)\,
S_b(p_b;Q^2,Q^2_0) \,
\left|{\cal M}_{ab\rightarrow cd}(Q^2)\right|^2 \,
T_c(p_c;Q^2,\mu^2_0)\,
T_d(p_d;Q^2,\mu^2_0),
\end{eqnarray}
where
${\cal M}$ stands for the lowest order QCD matrix element, and
$S_{\cdots}(p_{\cdots};Q^2,Q_0^2)$ is the Sudakov form factor for space-like
branchings \cite{webber}.
The latter one is operative only for primary partons which
did not yet participate in any interaction process. This factor evolves
the initialization energy scale $Q_0$ to the scale $Q$ of the first
interaction of the primary partons.
It accounts in such a way for the possible primary emissions of partons.
The outgoing particles are excited and can emit further partons. This
is described by the Sudakov form factor for time-like partons
$T_{\cdots}(p_{\cdots};Q^2,\mu_0^2)$,
where $\mu_0$ is a bound for the energy scale.
Below this scale perturbative QCD becomes unreliable and a
phenomenological hadronization scheme has to be employed.

In the PCM, Eq.~(1) is solved within a test-particle
formalism utilizing Monte Carlo techniques.
Besides the interaction kernel several quantum mechanical corrections
are implemented \cite{klaus_rep}.
Our calculations are based on the PCM version  VNI-2.0+.
This is a corrected version \cite{MB} of K. Geigers original code
VNI-2.0 \cite{klaus_rep}. The very recent PCM version VNI-3.1 \cite{klaus_new}
seems not yet to be sufficiently tested so that we prefer to employ our well
tested version VNI-2.0+.


\section{Production of SUSY particles in nucleon collisions}

Let us first consider the inclusive production of SUSY particles in
nucleon-nucleon collisions
\begin{eqnarray}
\label{process}
{\rm N} + {\rm N}' \rightarrow \tilde{\rm c} + \tilde{\rm d} + {\rm X},
\end{eqnarray}
where ${\rm N}$ and ${\rm N}'$ denote the two nucleons, $\tilde{\rm c}$
and $\tilde{\rm d}$ are two supersymmetric particles with masses
$m_{\tilde c, \tilde d}$, and ${\rm X}$ is
an arbitrary number of hadrons.
When taking only the first-chance collisions of partons,
the total cross section for the process (\ref{process})
can be calculated by
\begin{eqnarray}
\label{dyhaupt}
\sigma_{{\rm N N'}\rightarrow \tilde c\tilde d{\rm X}}
&=&
\frac 1s
\sum_{a,b=g,q,\bar q}\;
\int\limits_{s_{\rm min}}^s d\bar s
\int\limits_{{\hat s}/s}^1 dx \,
F^{\rm N}_{a}(x)\,F^{\rm N'}_{b}\left(\frac{\hat s}{sx}\right)
\frac1{x}
\,\sigma_{ab\tilde c\tilde d}(\hat s),
\end{eqnarray}
where $s_{\rm min}\equiv(m_{\tilde c}+m_{\tilde d})^2$,
and $F^{\rm N}_a(x)$ denotes the parton distribution function
of flavors $a$ of the nucleon $N$. We denote the cross section
(\ref{dyhaupt}) as Drell-Yan like cross section.
Later we shall compare this cross section with results of the PCM.
It is important to distinguish between the Mandelstam variables $s$
and $\hat s$: $s$ is always the squared invariant mass of the two
nucleons resp.\ nuclei, 
while $\hat s$ denotes the squared invariant mass of
an elementary parton-parton subprocess. We use the notion
$M^2 \equiv \hat s$.

In what follows we assume that the strong coupling constant
$\alpha_{\rm S}$ is the same for coupling to SM
particles as well as for coupling to SUSY particles.
The dependence of $\alpha_{\rm S}$ on the energy scale $Q$
is accounted for
up to the second order in $\ln(Q^2)^{-1}$
by \cite{pdg:96}
\begin{eqnarray}
\alpha_{\rm S}(Q^2)&=&
\frac{4\pi}{\beta_0\ln(Q^2/\Lambda^2)}
\left(1-\frac{2\beta_1}{\beta_0^2}
\frac{\ln\left[\ln(Q^2/\Lambda)\right]}{\ln(Q^2/\Lambda^2)}\right),
\end{eqnarray}
where $\beta_0 = 11- \frac 23 n_f$ and
$\beta_1 = 51-\frac19 3n_f$;
$n_f$ is the number of quark flavors with masses below $Q$.
If the energy scale $Q = \sqrt{\hat s}$ is chosen,
$\alpha_{\rm S}$ varies from $0.094$ to $0.067$
in the relevant region for $Q = 0.3 \ldots 14$ TeV.

The CTEQ4M parametrization \cite{cteq}
from CERN PDFLIB (version 7.07)
is employed for the parton distribution functions $F^{\rm N}_a$.

The various cross sections of reactions, where SUSY particles are involved,
can be found in Ref.~\cite{dawson:85}.
We recalculated and confirmed these results.

We assume that the $R$ parity is conserved in strong
interactions and that left and right chiral squarks display the same mass.
Precise values of masses are still lacking, of course.
Some actual mass limits are listed in Tab.~\ref{sqcd_part}
(cf.~\cite{Soff,pdg:96}).

Total cross sections for SUSY particle production are displayed
in Fig.~1 as a function of the available
center-of-mass system (CMS) energy $\sqrt{\hat s}$
of the incoming partons.
In Fig.~1 the mass for the lightest squark are taken as
$m_{\tilde q} = 230$ GeV and the gluino mass is
$m_{\tilde g} = 160$ GeV.
These masses are close to the lower limits for the case
$m_{\tilde g}\le m_{\tilde q}$.
The cross sections calculated with these values therefore serve
as upper limits.
One observes that the cross sections have their maximum slightly above the
threshold and then decrease as $\propto \hat s^{-1}$.
The reaction $ gg \to \tilde  g \tilde g$ has the largest cross section.
All the displayed cross sections depend sensitively on the unknown
SUSY particle masses. Below we discuss the dependence of the production rates
on these masses.

Fig.~2 depicts the individual cross sections for
SUSY particle production in proton-proton collisions in dependence on the
CMS energy $\sqrt{s}$ of the protons.
The figure covers the complete energy domain
$\sqrt{s} \le 14\,{\rm TeV}$ which is available at LHC.
The various combinations of outgoing strongly interacting SUSY particles
are separately displayed.
The gluino-gluino production dominates, followed by gluon-squark and
squark-anti-squark production.
The dominant parton processes are
$g g \to \tilde g \tilde g$,
$g q \to \tilde g\tilde q$,
and
$q_i\bar q_j \to \tilde q_i \tilde q_j$.

Fig.~3 shows the summed cross sections for the production of
$\tilde g \tilde g$, $\tilde g \tilde q$ and $\tilde q \tilde q$.
As to be expected from inspecting Fig.~2, the relation
$\sigma_{\tilde g \tilde g} > \sigma_{\tilde g \tilde q} >
\sigma_{\tilde q \tilde q}$ holds.

As mentioned above, the masses of the SUSY particles represent the
largest uncertainties since only lower bounds are known.
To illustrate the effect of a change of the SUSY particle masses we
depict in Fig.~4 the cross sections
for gluino-pair production in dependence on the gluino mass.
The mass of the lightest squark is set equal to the gluino mass, i.e.,
$m_{\tilde g}=m_{\tilde q}$, although the process
$ q\bar q\rightarrow \tilde g \tilde g $ can almost be neglected.
The gluino mass is chosen to be
$m_{\tilde g} =$ 50, 100, 250 and 500 GeV.
The cross sections for
$m_{\tilde g}=50, 100\,{\rm GeV}$
serve only for comparative purposes, since they are already ruled out
by the experiment.
If $\sqrt{s}$ is large the cross sections fall by more than one
order of magnitude when one doubles the gluino mass.
If the invariant mass fulfills $\sqrt{s} <$ 500 GeV,
the threshold energy $2m_{\tilde g}$ becomes important.
If $\sqrt{s}$ is ten times larger than the gluino mass,
the cross section is large enough to produce SUSY particles in
nucleon-nucleon collisions. Here we have assumed that a cross section
of $10^{-4}\,{\rm nb}$ is just
sufficient to be verified in experiments.
For gluino masses of 1 TeV and $\sqrt{s} =$ 14 TeV
the cross sections is indeed slightly above $10^{-4}\,{\rm nb}$.

The relation of the gluino mass to the lightest squark mass has also a strong
influence on the dominating elementary parton process.
To illustrate this effect we display in Fig.~5 the cross sections
for the case $m_{\tilde g}>m_{\tilde q}$
with $m_{\tilde g} =$ 220 GeV and
$m_{\tilde q} =$ 180 GeV. Again, the masses are just
above the actual lower boundaries for this case.
For $\sqrt{s} < $ 800 GeV the $\tilde q\tilde q^{(*)}$
production becomes dominant,
while above 7.5 TeV the
$\tilde g\tilde q$ production is most important.
The figure indicates that for $\sqrt{s}$ above
$14\,{\rm TeV}$ the $\tilde g\tilde g$ production
becomes the dominant process, as in Fig.~\ref{dyrates}.
This behavior follows from the form of the parton distribution
functions. The gluon distribution dominates the quark distributions,
but is concentrated at low momenta. The valence quark distributions
have contributions at significantly higher momenta.
Gluon-gluon interactions take place much more often than gluon-quark or
quark-quark interactions, since there are by far more gluons than
quarks. Otherwise, the CMS energy $\sqrt{\hat s}$
of the gluon-gluon interaction processes is much smaller.
Since the masses of supersymmetric particles are large, large
values of $\sqrt{\hat s}$ are needed.
Therefore, quark-quark interactions are much more
efficient for SUSY particle production than gluon-gluon interactions.
Gluon-quark interactions are somewhere in between gluon-gluon and
quark-quark interactions.


\section{Production of SUSY particles in nuclear collisions}

To calculate the SUSY particle production cross sections
in nuclear collisions,
we extrapolate appropriately the cross sections (\ref{dyhaupt}).
The difference between nucleons and protons can be neglected because
of the large energies which diminish the difference between $u$ and
$d$ quarks.
We consider symmetric heavy-ion collisions
where each nucleus consists of $A$ nucleons.
From parton cascade calculations (see below)
we know the dependence of the
nuclear cross section on the nucleon number,
\begin{eqnarray}
\label{dynuceq}
\sigma_{\rm tot}^{\rm AA}(s_{A})&=&
A^{4/3}\,\sigma_{\rm tot}^{\rm pp}(s_{A})\;,
\end{eqnarray}
where $\sqrt{s_A}$ is the CMS energy per nucleon of both nuclei.

Now we want to compare this SUSY particle production cross sections with
results of the PCM.
The cross sections for SUSY particle production are
relatively small compared to the
elementary QCD cross sections. It is therefore a good assumption that
the cascade will not be disturbed significantly if some SUSY particles
are produced.
$R$ parity conservation ensures that no created SUSY particle will be
lost by decay or interactions, though the original flavor may change.

The number $N_{\tilde c\tilde d}$ of produced pairs of SUSY particles
$\tilde c\tilde d$
can be estimated by
\begin{eqnarray}
\label{susyrate}
N_{\tilde c\tilde d}&=&\sum_{a,b}
\int\limits_0^\infty
dM\,
\frac{\hat\sigma_{ab\rightarrow\tilde c\tilde d}}{
\sum_{c,d}\hat\sigma_{ab\rightarrow cd}}
\sum_{c,d}\frac{dn_{ab\rightarrow cd}(M)}{dM},
\end{eqnarray}
where $\hat\sigma_{ab\rightarrow c d}$ is the total cross section of
an elementary two-parton process.
$dn_{ab\rightarrow cd}(M)/dM$ denotes the differential rate for
processes of the type $ab\rightarrow cd$. It is extracted from our
PCM calculation.
This quantity describes the distribution of binary parton collisions
as a function of $M \equiv \sqrt{\hat s}$.
For heavy particle production, the distribution at very large values of
$\sqrt{\hat s}$ is important. One intriguing question is now to investigate
whether the distribution changes with the size of the system.
It should be noted that at LHC the beam energies depend on the nuclei
to be accelerated since the beam energy per unit charge is constant.
This implies the following beam energies (in TeV) for various nuclei:
$^{16}$O, $^{32}$S: 3.5,
$^{56}$Fe: 3.25,
$^{208}$Pb: 2.76.

Fig.~7 shows the scaled production rate
$dn/dM$,
averaged over impact parameters and for different nuclei for LHC energies.
Since the cross section $\sigma$ is given by
\begin{eqnarray}
\label{nucsig}
\sigma&=&\pi b_{\rm max}^2\int dM\,\frac{dn}{dM}
\end{eqnarray}
we are going to depict the pure
rate $dn/dM$ multiplied by the factor $\pi b_{\rm max}^2$ .
The maximum impact parameter $b_{\rm max}$ is chosen to be twice
the radius of the colliding nuclei. Therefore, $b_{\rm max}$ depends on the
specific nuclei.

As seen in Fig.~7, the rate is rapidly increasing with growing nucleon number
$A$.
For given nucleon number $A$,
in the intermediate energy region (say 200 - 1000 MeV) the rate drops
roughly exponentially, and at $\sqrt{\hat s} >$ 1 TeV it flattens out.
In the case of large values of $M$ (say $ M >$ 1500 GeV),
the total number of interactions is in the order of unity. This
explains the large fluctuations in this region.
As mentioned above,
the values displayed in Fig.~7 are for fixed beam energies per charge unit.
In despite of the caused variation of beam energies, the flattening of the
rate $dn/dM$ sets in at $M \sim 1$ TeV. Limited statistics does not allow
to draw firm conclusions on a larger interval of accessible large-$M$
collisions in lead-lead reactions in comparison with pp reactions.
Nevertheless, it is fair to say, when inspecting Fig.~7, that in collisions
of heavy nuclei a similar range of $M$ is accessible as in collisions
of protons, even if $\sqrt{s}$ is only $\frac 13$ of the latter one.
The scaled rate at $M \sim 1$ TeV is in AA collisions by a factor of $10^4$
larger than in pp reactions.

For the calculation of SUSY particle production rates we need $dn/dM$
separately for each individual pair of initial particles. These rates
are depicted in Fig.~\ref{pbsplit}. Gluon-gluon and gluon-quark
processes occur most often, since the number of gluons
exceeds the number of quarks. For a large invariant masses
$M >$ 500 GeV the gluon-quark interactions dominate.
This corresponds to the effect discussed above, which is
caused by the large
momenta of the valence quarks. The number of quarks is not high
enough to lead to a significant contribution of quark-quark
interactions.
It is a remarkable result that $gq$ processes seems to be
most important for production of heavy particles in central nuclear
collisions.

Since we have calculated the distribution $dn/dM$ for various nuclei
we can proceed and calculate the dependence of the
interaction rate on the nucleon
number A. We have to integrate the distributions over $M$
weighted with $\pi b_{\rm max}^2$.
Fig.~\ref{nAfit} depicts the result in comparison with different fit
functions depending on the nucleon number $A$.
Obviously, the data points are
approximated best by a function of the type $a_0A^{4/3}$.
This result justifies our approach leading to Eq.~(\ref{dynuceq}).

Besides the cross sections for SUSY particles production, the total
cross sections for gluon and quark reactions of the SM
enter into  Eq.~(\ref{susyrate}).
Some of these cross section are infrared divergent and have to be
regularized.
For consistency, we employ the same regularization as in the PCM.
Fig.~\ref{smtot} depicts the regularized total
cross sections for various parton reactions in the relevant energy region.

By means of Eq.~(\ref{susyrate}) the SUSY production rate $N_{\tilde
c\tilde d}$ can be calculated.
The production cross section $\sigma_{\tilde c\tilde d}$
then yields
$\sigma_{\tilde c\tilde d} =
\pi b_{\rm max}^2\, N_{\tilde c\tilde d}$.
The results for two different mass scenarios of SUSY particles are listed in
Tab.~\ref{prodrat}.
The VNI-2.0+ parton cascade calculation is compared with the
Drell-Yan like rates at fixed beam energies per charge unit.
One reveals from Tab.~II two interesting features.
First, for proton-proton collisions,
VNI-2.0+ delivers cross sections which are at least by a factor
$\frac 13$ smaller than the Drell-Yan like rates.
Second, for nuclear collisions, VNI-2.0+ delivers at least
three times more SUSY particles than the extrapolated
Drell-Yan like rate predicts.
The latter one is the wanted collective effect.
When normalizing the VNI-2.0+ pp result
to the Drell-Yan like rate, the corresponding VNI-2.0+ results for
lead-lead collisions would be a factor 30 or more larger than the
extrapolated Drell-Yan like rate.
Even if one takes these numbers with some caution,
it appears that, due to secondary parton interactions, the high-energy
subprocesses are noticeably amplified in nuclear collisions.

To understand these effects better, one should keep in mind that in the
Drell-Yan like calculations only reactions between primary partons
(i.e., such ones which did not yet suffer an interaction)
are considered.
In the parton cascade calculations also the reactions with
secondary partons
(i.e., such ones which already suffer an interaction)
are taken into account. This leads generally to larger
rates. The effect is the stronger the heavier the nuclei are, since
the density grows with the number of nucleons per nucleus.
The energies of the secondary partons are lower compared to the primary
partons.
As stated above, the process $gq\rightarrow \tilde g\tilde q$
is the most dominant reaction for SUSY particle production for low
invariant masses just above the threshold.
Therefore, the process $gq\rightarrow \tilde g\tilde q$
becomes dominant if the secondary interactions become important.
This is to be contrasted with the Drell-Yan like rates (compare Tab.~II,
top rows) where the $\tilde g \tilde g$ channel for
$m_{\tilde g} > m_{\tilde q}$ dominates.

It can be observed (compare Tab.~II, bottom rows)
that $\tilde g\tilde q$ production
is the dominant process in
the case of $m_{\tilde g}>m_{\tilde q}$
in nucleon-nucleon calculations.
Extrapolated to nucleus-nucleus collisions,
$\tilde q\tilde q^{(*)}$ production becomes dominant.
This is due to the effect that the beam energy per nucleon is lower in
nucleus-nucleus collisions.
In this case, the higher energy of the quarks is
more important than the higher density of the gluons
(see also Fig.~\ref{dyrates2}).
Secondary interactions, however, change this feature and favor again the
$gq\rightarrow \tilde g\tilde q$ channel.

Finally let us comment on the different results of Drell-Yan like estimates
and VNI-2.0+ calculations for pp collisions.
Partially we can understand this
difference as a result of a non-perfect parton distribution
initialization in the PCM \cite{MB}, and too low phase space occupation
even in high-statistics runs.
Further, in the parton cascade it is rather likely that a
single parton with very high momentum can escape from the interaction
region in a proton-proton collision without any interaction at all.
This is supposed not to happen not so often
in collisions of heavy nuclei because of the
much higher particle density.
As a result, the PCM rates in proton-proton collision
are diminished. This effect is a consequence of the test-particle
character of the parton cascade calculation and seems to become severe
in extreme regions of the phase space.
As a test of our calculation routines we have used the fiducial masses
$m_{\tilde g} =$ 10 GeV and $m_{\tilde q} =$ 20 GeV and find that,
when summing over all channels, the Drell-Yan like rate and the
VNI-2.0+ result agree perfectly.

The actual production rate in the experiment is mainly determined by
the luminosity of the particle beams. Since the luminosities for
proton and lead beams at the LHC
will differ by a factor of about $10^7$, it
is very important to take the luminosity into account if one wants to
compare the production rates in proton-proton and lead-lead collisions.

Obviously, Tab.~\ref{prodrat} demonstrates clearly that one can not
expect to find signatures for SUSY particle production in collisions
of heavy nuclei easier than in proton-proton collisions.
The SUSY particle production rates in pp collisions are by a factor of
$10^3$ (or more) larger than in AA collisions.
The larger luminosity and the higher energy per nucleon in
proton-proton collisions provides a clear advantage in
the search of production of heavy particles.
In addition, it can be expected that it is much easier to analyze an
experimental event originating from a proton-proton collision
rather than
from nucleus-nucleus collision since much fewer particles are involved.

The following estimates rely on
the design data of LHC which envisage
a luminosity (in units of $10^{30}\,{\rm cm^{-2}s^{-1}}$)
of 0.002 of lead beams in contrast to $10^4$ for protons;
in addition, there will be two beam crossings of protons.
The best case scenario is to produce
$52$ pairs of SUSY particles per second and per beam
crossing in proton-proton collision with
masses $m_{\tilde g}=160\,{\rm GeV}$ and
$m_{\tilde q}=230\,{\rm GeV}$.
In lead-lead collision, only one pair will be produced in $145$
seconds.
Since it is probable that the masses are higher than assumed here, this can
be considered to represent an upper limit.
In the case of $m_{\tilde g} = m_{\tilde q} =$ 1 TeV,
it can be expected that one pair of SUSY particles will be produced
every $24$ minutes.
Although this seems to be a short period, it should be taken into
account that much more than one event is necessary to find a clear
evidence for SUSY production experimentally.


\section{Summary}

In summary
we estimate the production rates of supersymmetric particles in central
collisions of lead on lead at LHC energies within the parton cascade model
and compare with the production rates in pp collisions.
We do not find a sufficiently strong
cooperative effect in heavy-ion collisions which
would allow for more frequent production
of SUSY particles. Due to different beam energies and
luminosities for proton and lead beams,
it is therefore most likely that SUSY particles (provided they exist)
will be primarily found in proton-proton collisions at the LHC.
Even for such a very high masses of the gluino as
$m_{\tilde g} =$ 1 TeV
and of the lightest squark as $m_{\tilde q} =$ 1 TeV, it can be
expected that SUSY particles can be produced with a considerably high
rate to be detectable experimentally.

\subsection*{Acknowledgements}
One of the authors (M.W.B.) would like to thank the GSI Darmstadt, the BMBF,
the Land Sachsen and the DAAD for support; he also wants to thank the
Physics Faculty at Duke University for the warm hospitality.
The work is supported in part by BMBF grant 06DR829/1.


\newpage
\section{Tables}

\begin{table}
\begin{center}
\begin{tabular}{l|c|l}
SUSY particle &
mass limit $[{\rm GeV}]$ &
condition\\
\hline
squark $\tilde {\rm q}$ &
$ > 224$ &
$m_{\tilde g}\le m_{\tilde q}$\\
& $>176$ &
$m_{\tilde g}<300\,{\rm GeV}$\\
gluino $\tilde {\rm g}$ &
$>154$ & $m_{\tilde g}< m_{\tilde q}$\\
& $>212$ &
$m_{\tilde g}\ge m_{\tilde q}$\\
\end{tabular}
\end{center}
\caption{
Strongly interacting supersymmetric particles and their mass limits.
$\tilde q$ denotes the lightest squark.
The mass limits depend on the mass relation of the gluino to the
lightest squark.}
\label{sqcd_part}
\end{table}

\vspace*{1cm}

\begin{table}
\renewcommand{\arraystretch}{1.5}
\begin{center}
\begin{tabular}{l|l|c||d|d|d|d}
&&  & \multicolumn{2}{c|}{Drell-Yan like} &
\multicolumn{2}{c|}{VNI-2.0+}  \\ \cline{3-7}
SUSY masses &
& SUSY pair
& ${\rm p}$ & $\vphantom{P}^{208}{\rm Pb}$&
 p & $\vphantom{P}^{208}{\rm Pb}$ \\ \hline\hline
&
&$\tilde g\tilde g$& 3.68 & 345 & 0.365 & 1030 \\
&$\sigma_{\rm tot}\,[{\rm nb}]$&$\tilde g\tilde q$&
 1.14 & 153 & 0.374 & 2000 \\
$m_{\tilde g}=160\,{\rm GeV}$ &
& $\tilde q\tilde q^{(*)}$ & 0.367 & 66.9 & 0.0309 & 425 \\ \cline{2-7}
$m_{\tilde q}=230\,{\rm GeV}$ &
&$\tilde g\tilde g$& 36.8 & 6.90${}\times10^{-4}$ & 3.65 &
0.00206 \\
& Rate $[{\rm s}^{-1}]$
&$\tilde g\tilde q$&
 11.4 & 3.06${}\times10^{-4}$ & 3.74 & 0.004 \\
& & $\tilde q\tilde q^{(*)}$ & 3.67 & 1.34${}\times10^{-4}$ & 0.309 &
8.5${}\times10^{-4}$ \\ \hline
&
&$\tilde g\tilde g$& 0.797 & 56.2 & 0.119 & 338 \\
&$\sigma_{\rm tot}\,[{\rm nb}]$&$\tilde g\tilde q$&
 0.852 & 112 & 0.298 & 1580 \\
$m_{\tilde g}=220\,{\rm GeV}$ &
& $\tilde q\tilde q^{(*)}$ & 0.699 & 140 & 0.0455 & 688 \\ \cline{2-7}
$m_{\tilde q}=180\,{\rm GeV}$ &
&$\tilde g\tilde g$& 7.97 & 1.12${}\times10^{-4}$ & 1.19 &
6.76${}\times10^{-4}$ \\
& Rate $[{\rm s}^{-1}]$
&$\tilde g\tilde q$&
 8.52 & 2.24${}\times10^{-4}$ & 2.98 & 0.00316 \\
& & $\tilde q\tilde q^{(*)}$ & 6.99 & 2.80${}\times10^{-4}$ & 0.455
& 0.00138
\end{tabular}
\end{center}
\caption{
Cross sections and production rates for SUSY particle production for
different masses of the gluino and the lightest squark.}
\label{prodrat}
\end{table}


\section{Figures}

\begin{figure}
\centering
\leavevmode
\epsfxsize=\textwidth
\epsffile{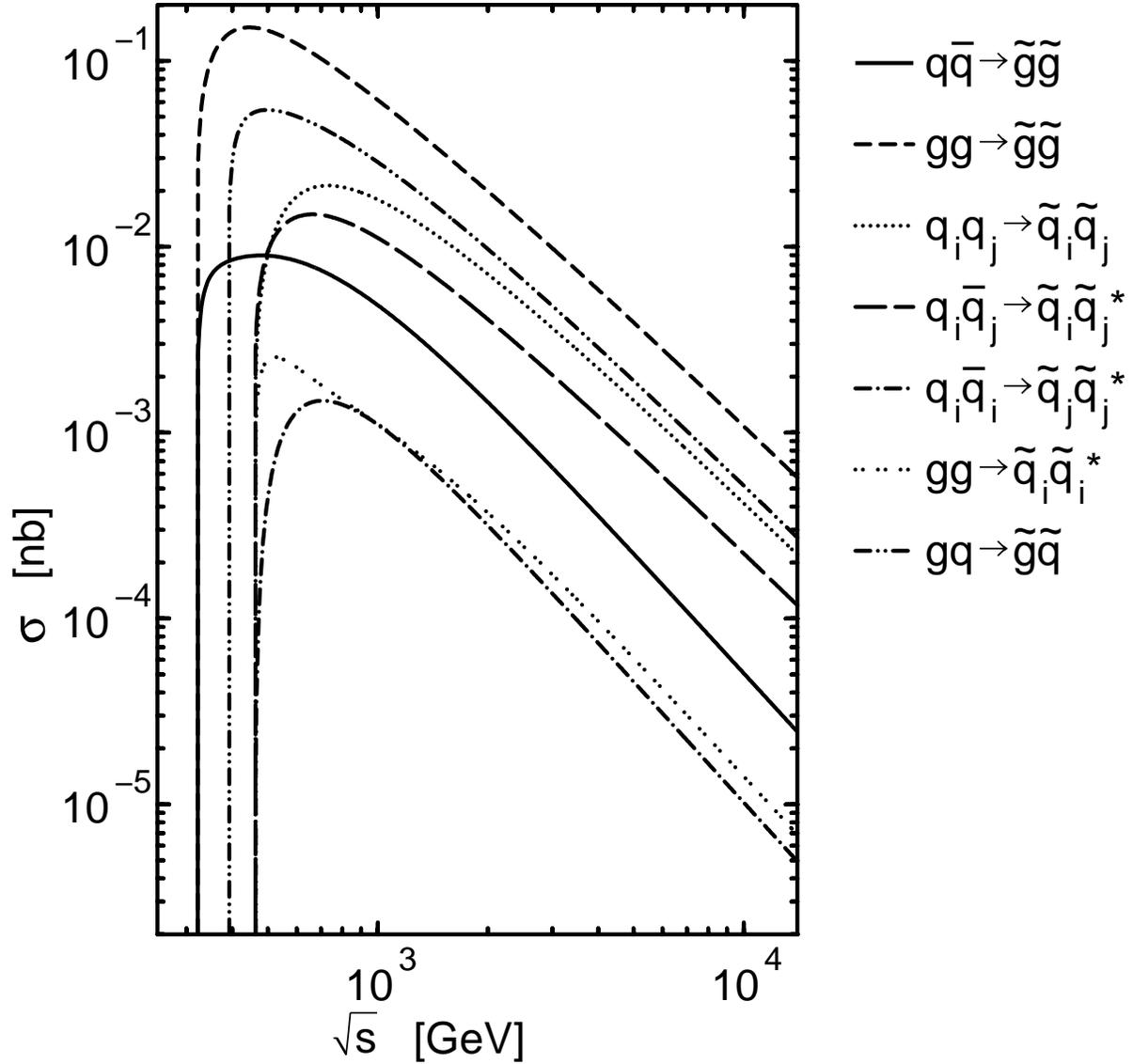}
\caption{
Total cross sections of seven fundamental SUSY particle
production processes as a function of the CMS energy
$\protect\sqrt{\hat s}$.
The masses are chosen to be
$m_{\protect\tilde q} =$ 230 GeV for the lightest squark, and
$m_{\protect\tilde g} =$ 160 GeV for the gluino.}
\label{txs}
\end{figure}

\begin{figure}
\centering
\leavevmode
\epsfysize=\textwidth
\epsffile{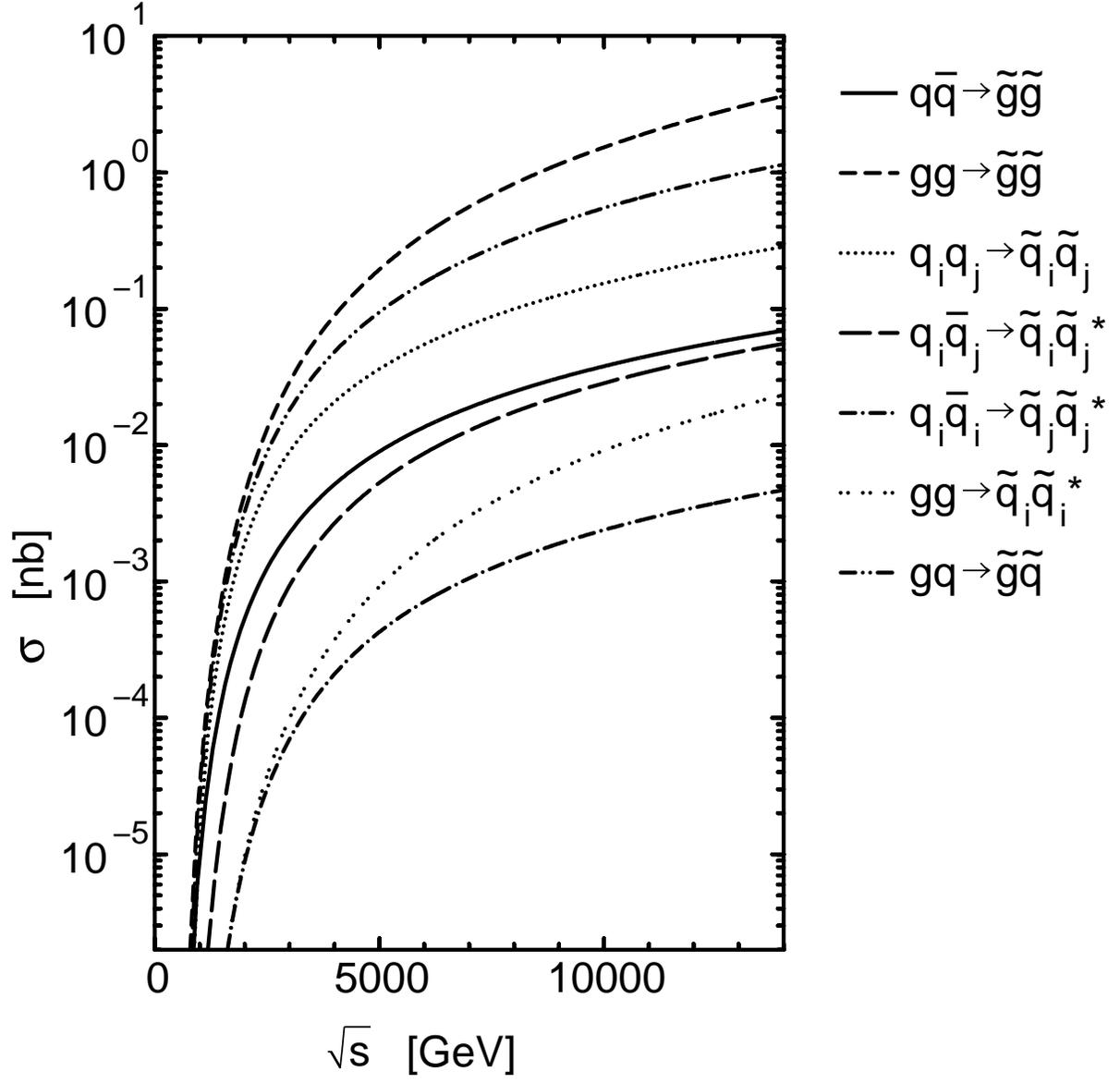}
\caption{Drell-Yan like production of SUSY particles in proton-proton
collisions. The cross sections of individual subprocesses are
depicted separately.
The masses of the SUSY particles are
$m_{\protect\tilde q} =$ 230 GeV and
$m_{\protect\tilde g} =$ 160 GeV.}
\label{dytot}
\end{figure}

\begin{figure}
\centering
\leavevmode
\epsfysize=\textwidth
\epsffile{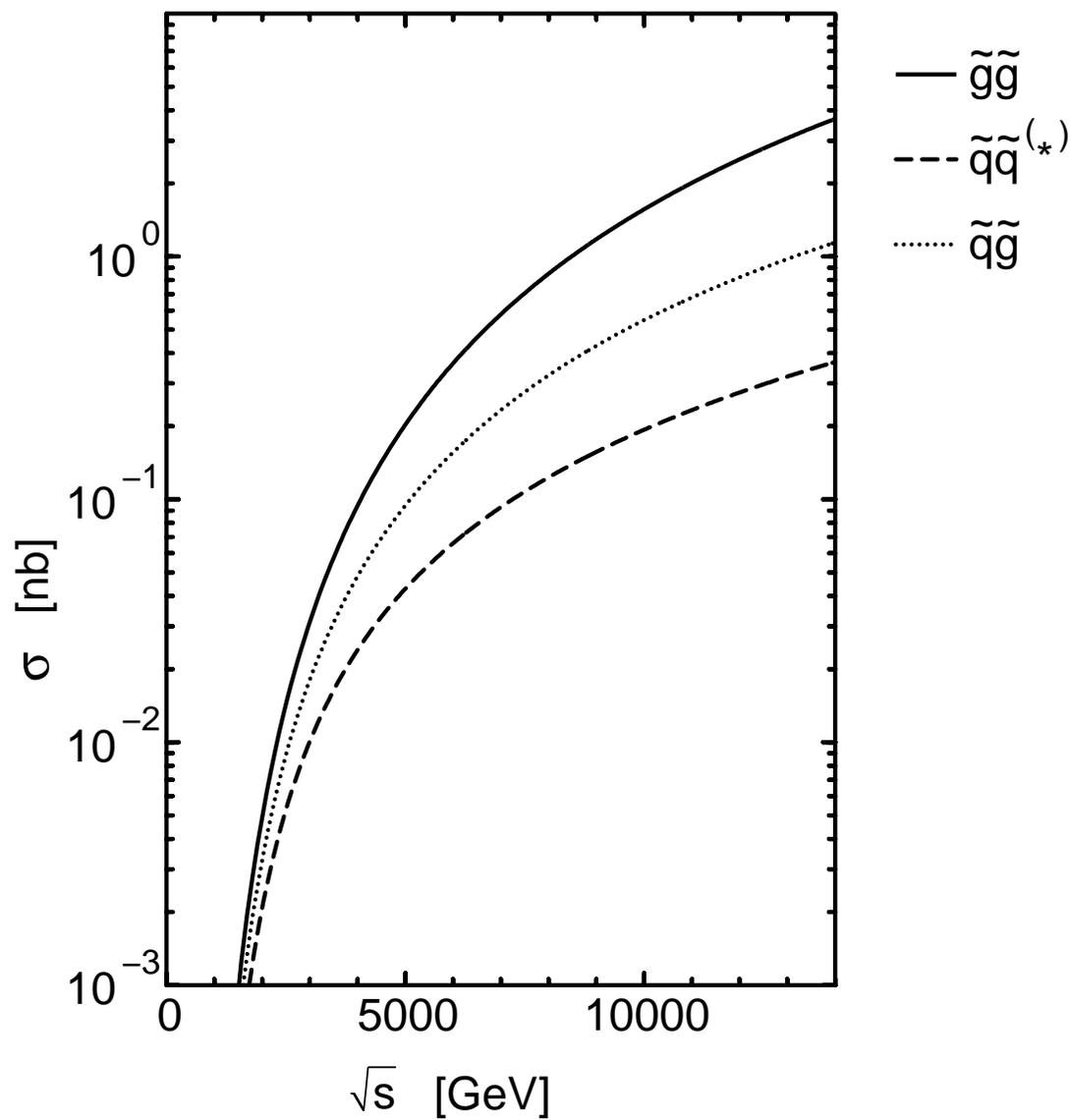}
\caption{
As in fig.~\protect\ref{dytot},
but summed over all processes with the same final
SUSY particle pair.}
\label{dyrates}
\end{figure}

\begin{figure}
\centering
\leavevmode
\epsfysize=\textwidth
\epsffile{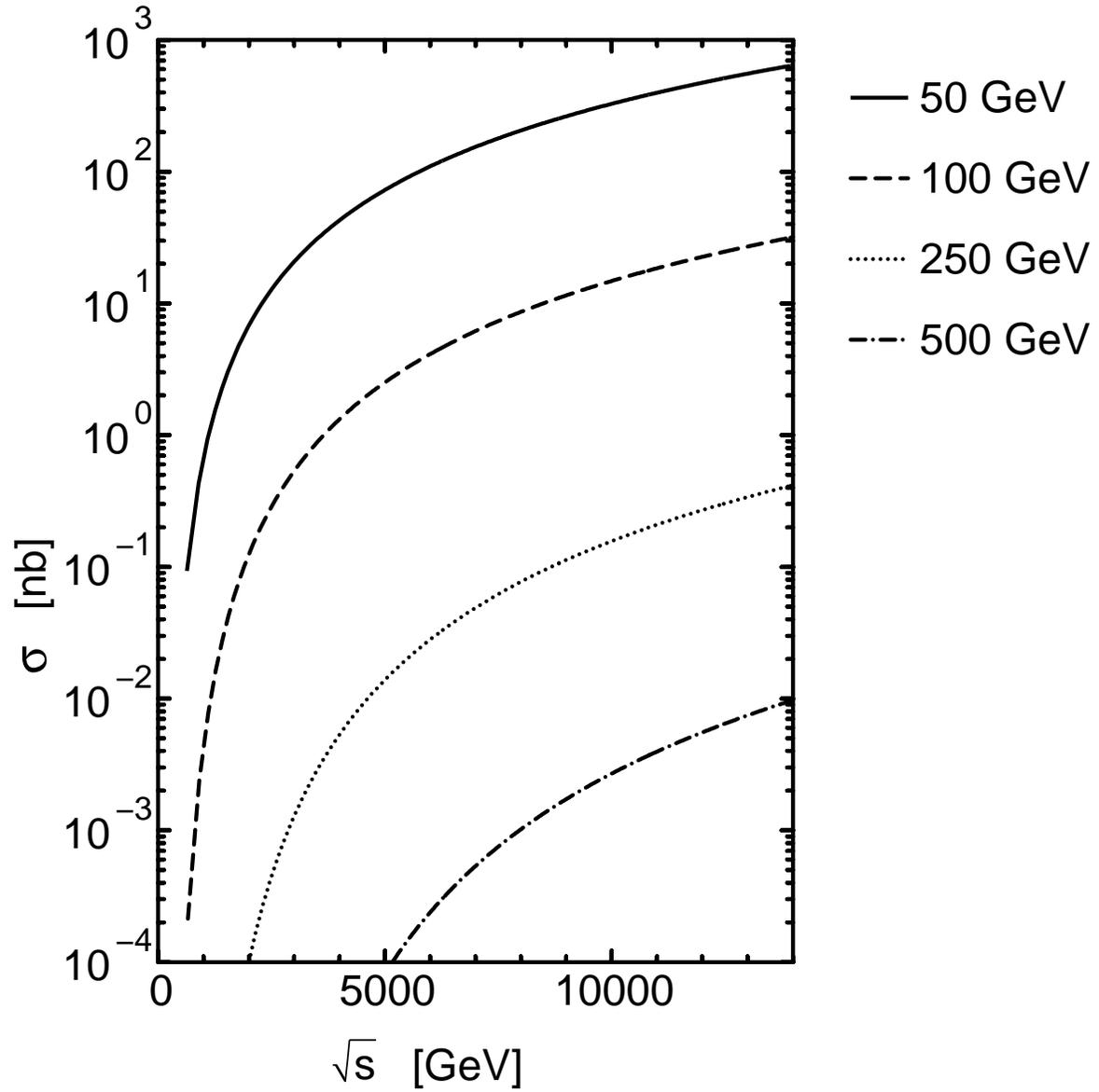}
\caption{Drell-Yan like production of gluino pairs in proton-proton
collisions for various assumptions on the gluino mass.
We use $\protect m_{\protect\tilde g} = \protect m_{\protect\tilde q}$.}
\label{dygg}
\end{figure}

\begin{figure}
\centering
\leavevmode
\epsfysize=\textwidth
\epsffile{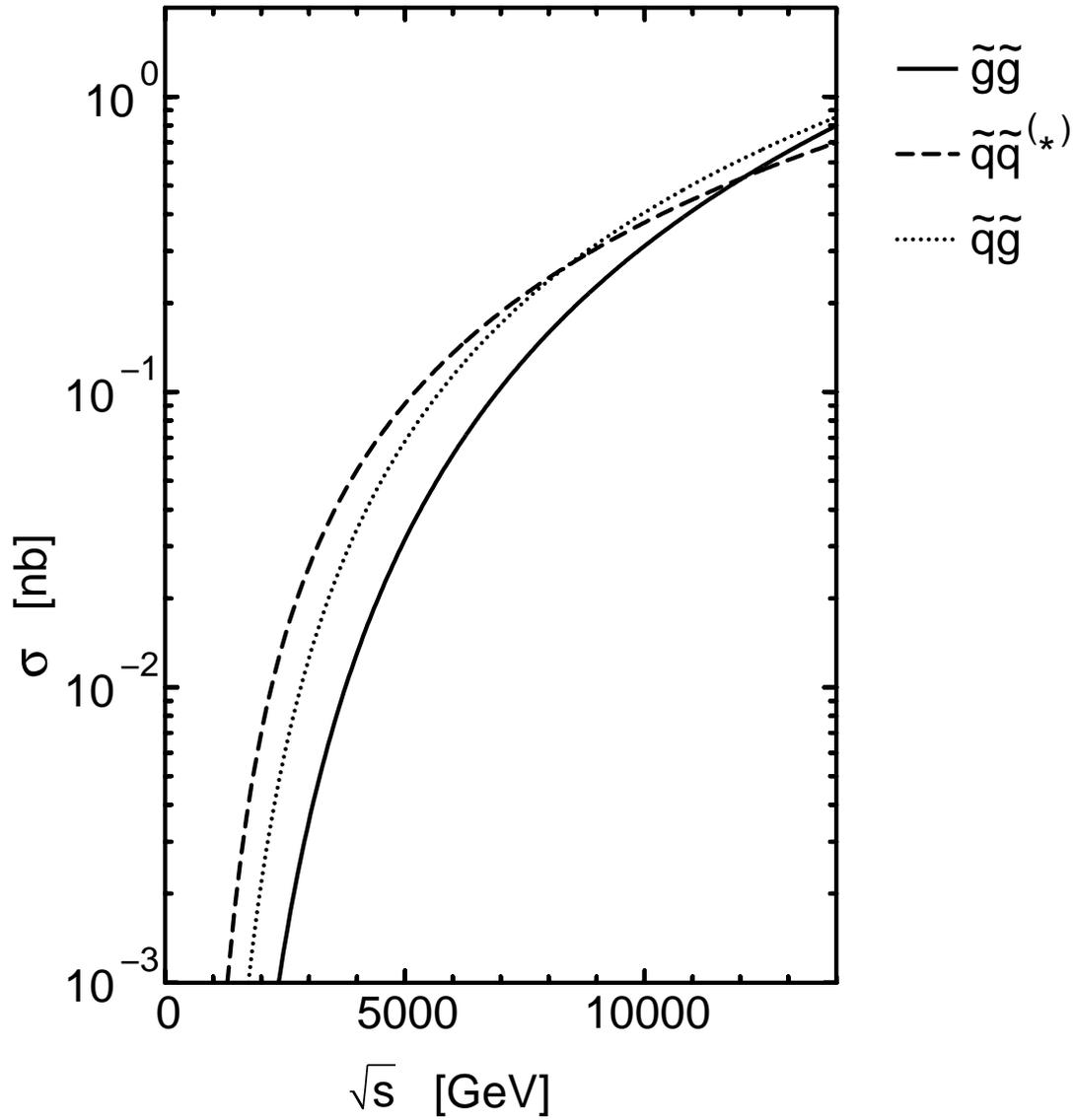}
\caption{
The same as in Fig.~\protect\ref{dyrates} for the case
$\protect m_{\protect\tilde g} > m_{\protect\tilde q}$.
The lightest squark mass is
$\protect m_{\tilde q} =$ 180 GeV,
while the gluino mass is $m_{\protect\tilde g} =$ 220 GeV.}
\label{dyrates2}
\end{figure}

\begin{figure}
\centering
\leavevmode
\epsfysize=\textwidth
\epsffile{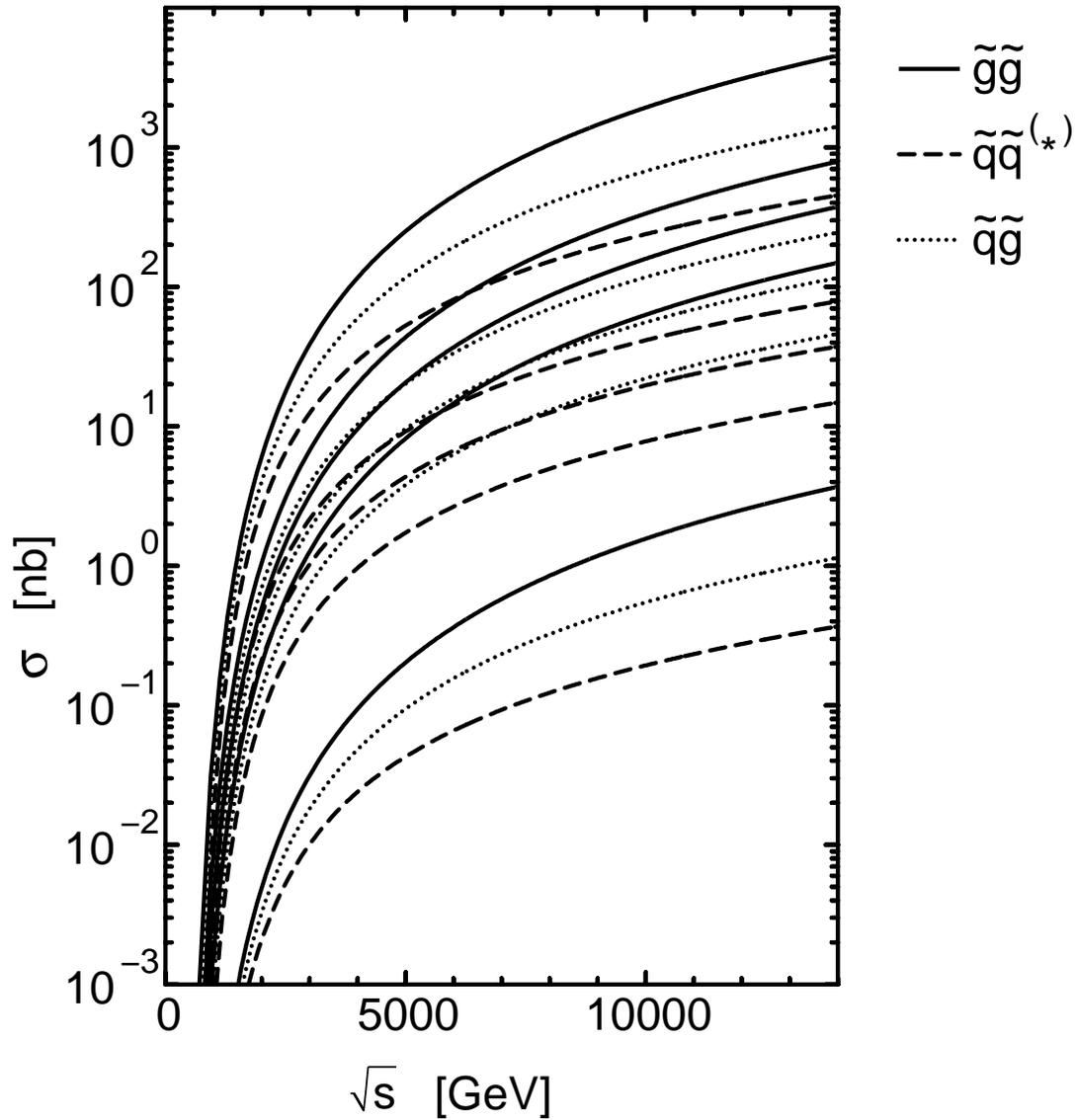}
\caption{
Extrapolated Drell-Yan like production cross sections of SUSY particles in
nucleus-nucleus collisions for different nuclei.
The masses of the SUSY particles are identical with those
in Fig.~\protect\ref{dyrates}.
From bottom to top, the cross sections are shown for
protons, $\protect\vphantom{V}^{16}{\rm O}$,
$\protect\vphantom{V}^{32}{\rm S}$,
$\protect\vphantom{V}^{56}{\rm Fe}$ and
$\protect\vphantom{V}^{208}{\rm Pb}$.}
\label{dynuc}
\end{figure}

\begin{figure}
\centering
\leavevmode
\epsfysize=\textwidth
\epsffile{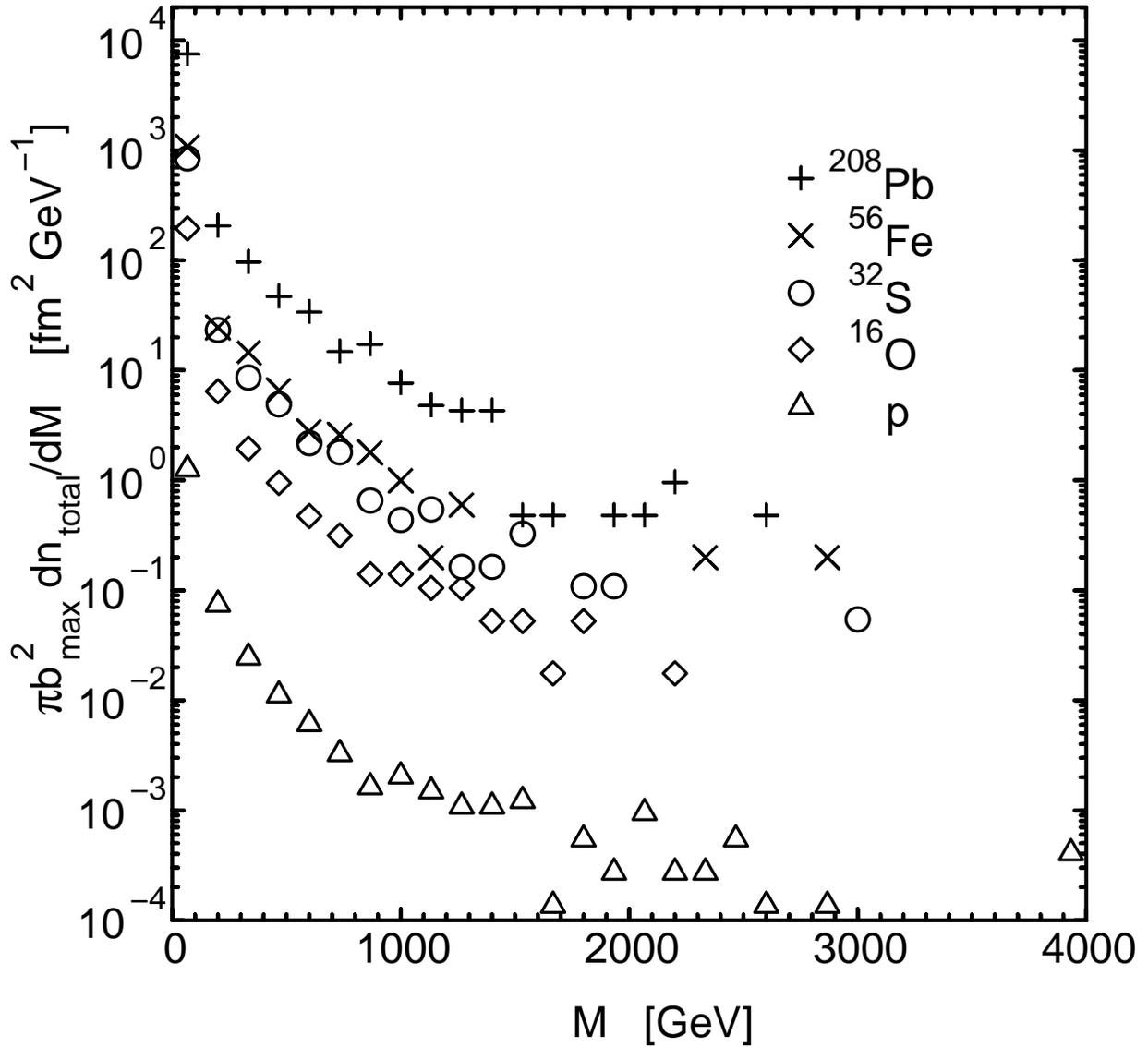}
\caption{
The total $\protect\pi b^2_{\protect \rm max} \, dn/dM$ distribution
for different nuclei at
7 TeV per charge unit in dependence on the invariant mass
$\protect M \protect\equiv \hat s^{1/2}$ 
of the parton-parton process. 
$\protect b_{\protect \rm max}$ is the maximum impact parameter.
The distributions were obtained by averaging over several collisions.
The fluctuations in the high $\protect M$ region are due to the 
low statistics in the high energy tail of the distributions.}
\label{nofA}
\end{figure}

\begin{figure}
\centering
\leavevmode
\epsfysize=\textwidth
\epsffile{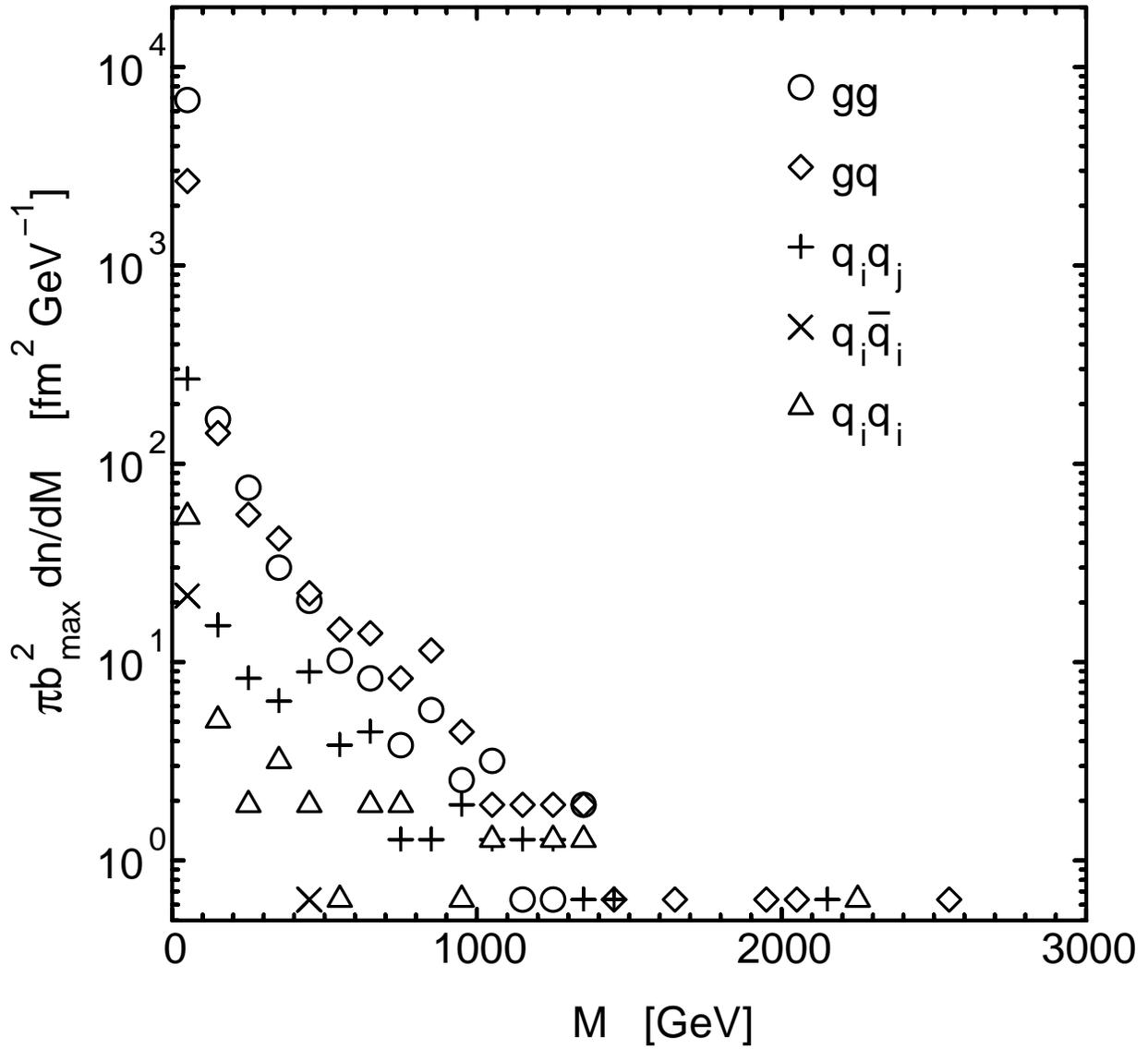}
\caption{
The $dn/dM$ distribution for
lead-lead collisions
at beam energies of 2.76 ATeV.
The $dn/dM$ distributions are depicted separately for
different combinations of colliding initial partons.}
\label{pbsplit}
\end{figure}


\begin{figure}
\centering
\leavevmode
\epsfysize=\textwidth
\epsffile{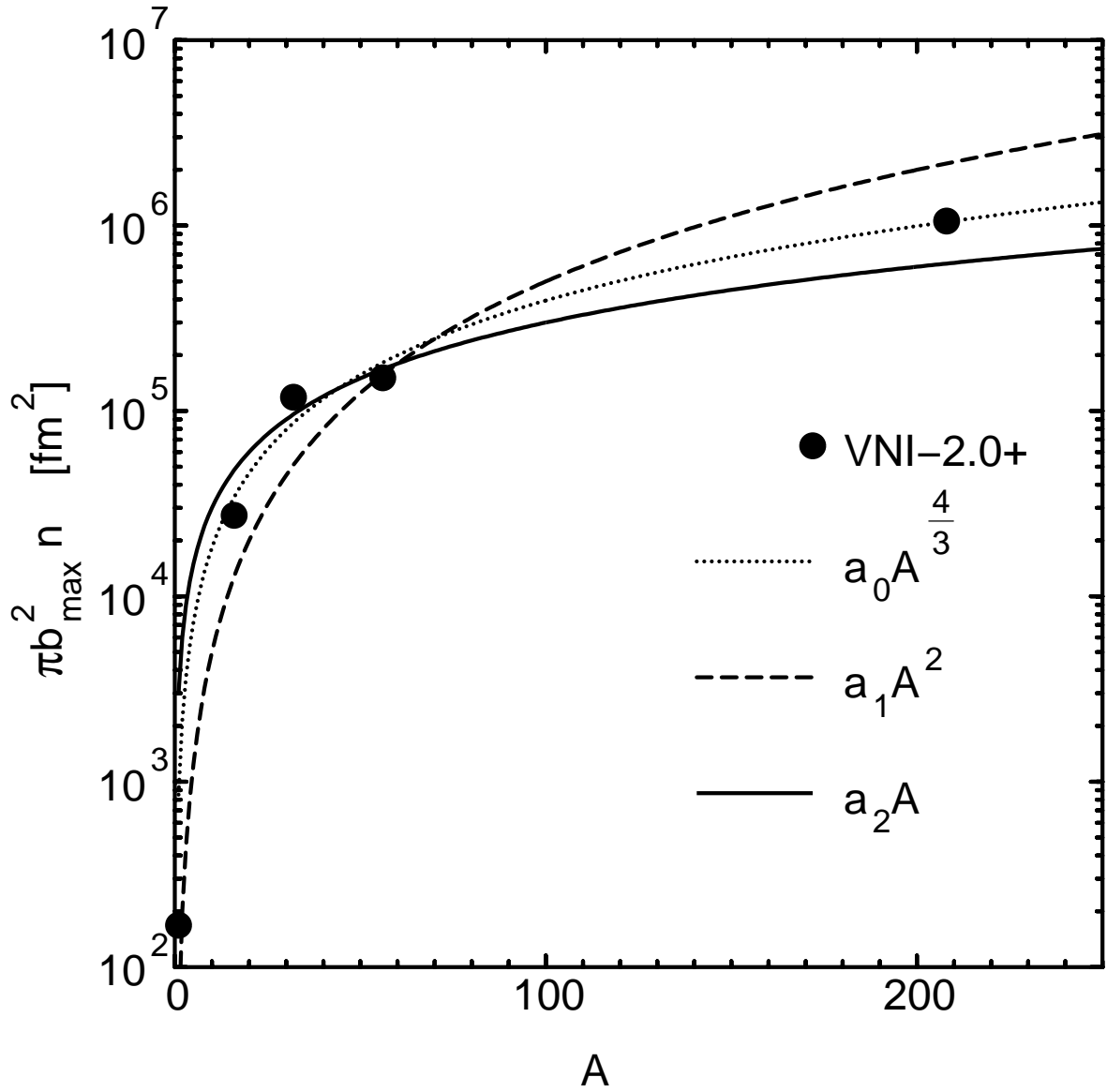}
\caption{
The integrated binary collision rate as a function of the
nucleon number $A$ compared to various fit functions.}
\label{nAfit}
\end{figure}

\begin{figure}
\centering
\leavevmode
\epsfxsize=\textwidth
\epsffile{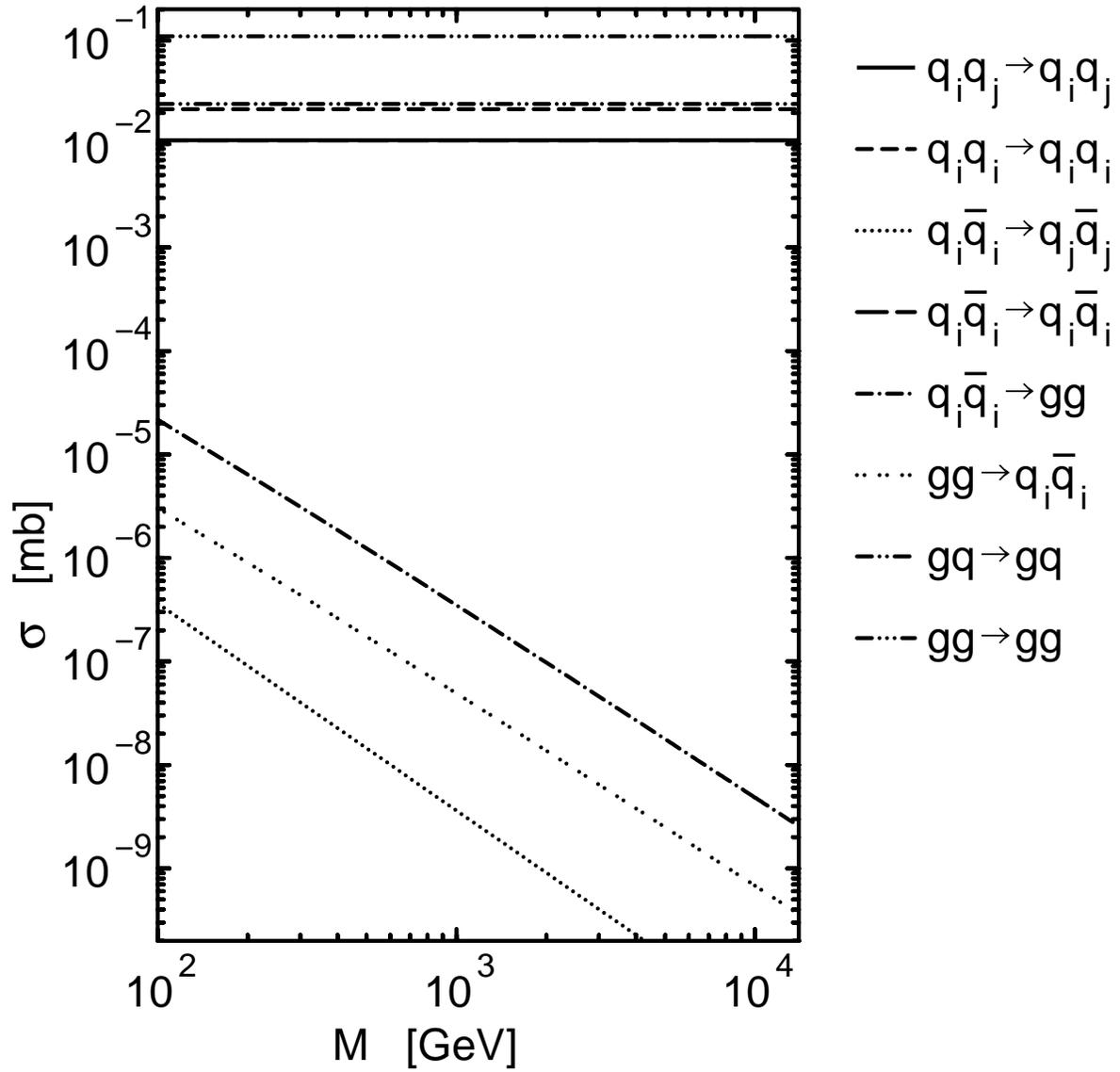}
\caption{
The total cross sections of the fundamental QCD procesess for
massless quarks and gluons.}
\label{smtot}
\end{figure}

\end{document}